\documentclass[aps,prb,superscriptaddress,floatfix,twocolumn,showpacs,amssymb]{revtex4}

\usepackage{amsmath}%
\usepackage{amssymb}
\usepackage[dvips]{graphicx}
\usepackage{bm}
\usepackage{latexsym}
\usepackage{color}
\usepackage[hypertex]{hyperref}
\usepackage{ulem}

\begin{document}
\title{Conformal invariance, multifractality, and finite-size scaling
at Anderson localization transitions in two dimensions}

\author{H. Obuse}

\affiliation{Condensed Matter Theory Laboratory, RIKEN, Wako,
Saitama 351-0198, Japan}
\affiliation{Department of Physics, Kyoto University, 060-8502 Kyoto, Japan}
\affiliation{James Franck Institute, University of Chicago, 5640
South Ellis Avenue, Chicago, Illinois 60637, USA}
\author{A. R. Subramaniam}
\altaffiliation[Present address: ]{FAS Center for Systems Biology,
Harvard University, Cambridge, Massachusetts 02138, USA.}
\affiliation{James Franck Institute, University of Chicago, 5640
South Ellis Avenue, Chicago, Illinois 60637, USA}
\author{A. Furusaki}
\affiliation{Condensed Matter Theory Laboratory, RIKEN, Wako,
Saitama 351-0198, Japan}
\author{I. A. Gruzberg}
\affiliation{James Franck Institute, University of Chicago, 5640
South Ellis Avenue, Chicago, Illinois 60637, USA}
\author{A. W. W. Ludwig}
\affiliation{Department of Physics, University of California, Santa
Barbara, California 93106, USA}

\begin{abstract}

We generalize universal relations between the multifractal
exponent $\alpha_0$ for the scaling of the typical wave
function magnitude at a (Anderson) localization-delocalization
transition in two dimensions and the corresponding
critical finite size
scaling (FSS) amplitude $\Lambda_c$ of the typical localization
length in quasi-one-dimensional (Q1D) geometry: (i) When
 \textit{open} boundary conditions are imposed in the transverse
direction of Q1D samples (strip geometry), we show that the
corresponding
critical FSS amplitude $\Lambda_c^o$ is universally
related to the \textit{boundary} multifractal exponent
$\alpha_0^s$ for the typical wave function amplitude along a
straight boundary (surface).
(ii) We further propose a
generalization of these universal relations
to those symmetry classes whose density of states
vanishes at the transition.
(iii) We verify our generalized relations
[Eqs.~(\ref{eq:ours-bulk}) and
(\ref{eq:ours-surface})] numerically for the following four types
of two-dimensional Anderson transitions: (a) the
metal-to-(ordinary insulator) transition in the spin-orbit (symplectic)
symmetry class,
(b) the metal-to-($\mathbb{Z}_2$ topological insulator)
transition which is also in the spin-orbit (symplectic) class,
(c) the integer quantum Hall plateau transition, and (d)
the spin quantum Hall plateau transition.
\end{abstract}


\date{May 23, 2010}

\maketitle

\section{Introduction}

Localization-delocalization (LD) or Anderson localization
transitions of non-interacting electrons are continuous phase
transitions driven by disorder.\cite{anderson, Wegner,
LeeRamakrishnan, Efetov, evers_review} When disorder is weak,
the single-electron wave functions are extended over the whole
sample.
Sufficiently strong disorder localizes electrons
within a finite region in space. The linear size of this region
is the localization length $\xi$ characterizing the typical
size of the wave functions $\psi(\bm{r})$.\cite{anderson} As
the disorder strength is reduced, the localization length
increases and eventually diverges at an LD transition point.
The localization length is the analogue of the correlation
length at  non-random continuous phase transitions. At the LD
transition point, wave function amplitudes obey
scale-invariant, multifractal
statistics;\cite{CastellaniPeliti, WegnerMultifractal,
janssen_review, mirlin_review} that is, the disorder-averaged
$q$-th moment of the square of the absolute value of wave
function has a power-law dependence on the linear dimension $L$
of the system, with an exponent that is a non-linear function
of $q$.\cite{CastellaniPeliti, WegnerMultifractal,
evers_review}

Let us recall that continuous phase transitions in
{\it non-random} systems are known to be quite generally
described by conformally-invariant field theories.
Conformal symmetry is especially powerful in two dimensions
(2D), where its presence
leads to an infinite number of symmetry constraints. This, in many
cases, allows for a rather complete
description of critical properties.\cite{BPZ, yellowbook}
Effective (field) theories describing
the {\it random}
LD transitions are also expected
to possess conformal symmetry.
In fact, we have recently shown by numerical
simulations of a standard LD transition occurring in two dimensions,
namely of the metal-insulator transition in the
2D spin-orbit (symplectic) symmetry class,\cite{SpinOrbitRef}
that multifractal exponents of critical wave functions evaluated on
a straight boundary and those at a corner are related through a
simple relation dictated by conformal symmetry.\cite{obuse2007}

Conformal symmetry is known to impose strong constraints on
finite-size scaling (FSS) for phase transitions in
{\it non-random}
systems with quasi-one-dimensional (Q1D) geometry.
For these systems Cardy has shown\cite{cardy1984}
that the correlation length $\xi$ which
characterizes the decay of the two-point correlation function
of any (conformal primary\cite{BPZ,ExplainPrimary}) operator
along a cylinder or a strip of width $M$, is
related to the bulk ($x_b$) or surface ($x_s$) scaling
dimension of the operator in two dimensions through
\begin{equation}
\frac{M}{\xi}=\left\{
\begin{array}{ll}
2\pi x_b, & \mbox{cylinder \ (periodic BC)},\\
\pi x_s, & \mbox{strip \ (open BC)}.
\end{array}
\right.
\label{eq:cardy}
\end{equation}
Here BC stands for boundary conditions imposed in the
transverse direction.

The generalization of Eq.\ (\ref{eq:cardy}) to scale-invariant
disordered 2D systems was first provided in the study of random
2D diluted ferromagnets in Ref.\ [\onlinecite{Ludwig1990}] [for
the 2D bulk exponents and Q1D cylinder geometry (periodic
BCs)]. In a random system the scaling of an observable (such
as, for example, a ``spin'') is in general characterized by the
set of scaling dimensions $x_q$ of all its $q$-th moment
disorder averages. Equation (\ref{eq:cardy})
generalizes\cite{Ludwig1990} to all these moments. In
particular, the correlation length $\xi_q$ characterizing the
exponential decay of the $q$-th moment of a correlation
function of the observable in Q1D cylinder geometry is related
to the 2D scaling exponent by
\begin{equation}
\frac{M}{\xi_q}=
2\pi x_q, \qquad {\rm cylinder \ (periodic\  BC)}.
\label{eq:ludwig-random}
\end{equation}
At the same time, by using an expansion about $q=0$ of the $q$-th moments
in the 2D system and in the Q1D cylinder geometry,
it was demonstrated
in Ref.\ [\onlinecite{Ludwig1990}] that
such a relationship holds
also for the corresponding
``{\it typical}'' quantities referring to a fixed disorder realization.
In particular, if $\alpha_0$ denotes the
{\it typical}\cite{Ludwig1990} 2D bulk scaling dimension
of the observable, and if
$1/\xi$ denotes the Lyapunov
exponent characterizing the inverse of the {\it typical} Q1D
correlation length in cylinder geometry,
then again the relationship
\begin{equation}
\frac{M}{\xi}=
2\pi \alpha_0, \qquad {\rm cylinder \ (periodic \  BC)}
\label{eq:ludwig-random-Lyapunov}
\end{equation}
holds.

\begin{table*}[t]
\caption{A list of $\alpha_0$, $x_\rho$ and $\Lambda_c$ for
various universality classes. The values of $\alpha_0$ marked
by $*$ are from the references listed in the last column. Those
without $*$ are obtained in this paper. The fifth column shows
$\Lambda_c$ calculated from $\alpha_0^b$ using Eq.\
(\ref{eq:ours-bulk}) and from $\alpha_0^s$ using Eq.\
(\ref{eq:ours-surface}), combined with Eq.\ (\ref{DEFLambdac}).
These values of $\Lambda_c$ should be compared with those
obtained from fitting (as explained in Section III) and shown
in the sixth column. } \label{tab:alpha_0}
\centering
\begin{tabular}{lcccccc}
\hline
\hline
system & BCs &$\alpha_0$ & $x_\rho$ & $\Lambda_c$ from $\alpha_0$
& $\Lambda_c$ from fit & Ref.\\
\hline
symplectic (M-I) & open & $2.429\pm0.006$ & $0$ &
$1.48\pm0.02$ & $1.50\pm0.01$ & this paper \\
symplectic (M-QSH) & open &$2.091\pm0.002^{*}$ & $0$ &
$7.00\pm0.15$ & $7.20\pm0.01$ &[\onlinecite{shinsei09}]\\
IQH & open &$2.385\pm0.003$ & $0$ &
$1.654\pm0.013$  & $1.624\pm0.002$ & this paper \\
SQH in class C & periodic & $2.137^{*}$ & $1/4$ &
$0.8225$  & $0.8189\pm0.0004$ & [\onlinecite{evers03}]\\
SQH in class C & open & $2.326^{*}$ & $1/4$ &
$1.105$ & $1.101\pm0.002$ &[\onlinecite{subramaniam}]\\
\hline
\hline
\end{tabular}
\end{table*}

Later, Refs.~[\onlinecite{janssen_review,dohmen96}] proposed a
corresponding formula in the context of LD transitions in two
dimensions,
\begin{equation}
\frac{M}{\xi_p} = 2\pi(\alpha_0^b-2),
\label{eq:janssen}
\end{equation}
[the shift by two between  the r.h.s.\ of Eq.\ (\ref{eq:janssen})
and of Eq.\ (\ref{eq:ludwig-random-Lyapunov}) arises from
different conventions].
Here, $\xi_p$ is the typical Q1D localization length in
\textit{cylinder} geometry (the subscript $p$ of $\xi_p$ denotes
periodic BCs imposed in the transverse direction). The exponent
$\alpha_0^b$ in Eq.\ (\ref{eq:janssen}) characterizes the
scaling of a typical critical wave function amplitude in the
\textit{bulk} of a 2D system of linear dimension $R$,
\begin{equation}
\overline{\ln|\psi(\bm{r})|^2}\sim -\alpha_0^b\ln R,
\label{eq:alpha_0}
\end{equation}
where the overbar stands for the disorder average. Equation
(\ref{eq:janssen}) has been confirmed numerically for the
integer quantum Hall (IQH) plateau
transition\cite{janssen_review, dohmen96, evers01} and for the
2D metal-insulator transition in the spin-orbit (symplectic)
symmetry class.\cite{merkt98, obuse2007, mildenberger07}

We note that the relation (\ref{eq:janssen}), in the form
presented, is only valid for systems in which the average bulk
density of states (DOS) is constant and non-vanishing at the
transition. This is the case for LD transitions in the three
Wigner-Dyson classes. These include the IQH plateau transition
and the LD transition in the spin-orbit (symplectic) class.
However, as is now well known, there are symmetry classes in
which the DOS vanishes at the transition. This is the case, for
example, for the so-called spin quantum Hall transition 
of the Bogoliubov-de Gennes (BdG) quasiparticles
in
symmetry class C\cite{gruzberg99, beamond02, mirlin03}
(in the nomenclature of Ref. \onlinecite{AltlandZirnbauer}).

In this paper we derive a generalization of the relationship
(\ref{eq:janssen}) between the exponent $\alpha_0^b$ and the
typical Q1D correlation length ${\xi_p}$ for LD transitions
in 2D with a {\it vanishing} critical DOS. The result is
\begin{align}
\frac{M}{\xi_p} = 2\pi (\alpha_0^b - 2 + x_\rho),
\label{eq:ours-bulk}
\end{align}
where the exponent $x_\rho$ characterizes the critical behavior
of the (bulk) DOS ($x_\rho = 0$ in the Wigner-Dyson classes).

Furthermore, we derive a FSS formula for the typical Q1D
localization length, when \textit{open} BCs are imposed in the
transverse direction. The specific open BC we consider in this
paper is a reflecting BC which means that the system simply
ends at the boundary, so that there is no current flowing
across the boundary. The second line of Eq.\ (\ref{eq:cardy})
suggests that the localization length should be related to a
surface exponent characterizing multifractality of critical
wave functions near boundaries of disordered
systems.\cite{subramaniam} Indeed, our result is the formula
\begin{equation}
\frac{M}{\xi_o} = \pi(\alpha_0^s - 2 + x_\rho),
\label{eq:ours-surface}
\end{equation}
where now $\alpha_0^s$ is the surface (i.e., boundary) exponent
characterizing scaling of a typical wave function amplitude
near a straight (reflecting) boundary. $\alpha_0^s$ is defined
in the same way as $\alpha_0^b$ in Eq.\ (\ref{eq:alpha_0}),
except that now the point $\bm{r}$ is close to a straight
boundary of the 2D system of linear dimension $R$. The typical
Q1D localization length $\xi_o$ is computed in the geometry
of a strip of width $M$ with open (reflecting) BCs imposed in
the transverse direction (the subscript $o$ stands for
``open'').

The organization of this paper is as follows. In Sec.\ II we
derive Eqs.\ (\ref{eq:ours-bulk}) and (\ref{eq:ours-surface}).
In Sec.\ III we verify both these equations numerically by
computing the critical FSS amplitude ($\Lambda_p$ or
$\Lambda_o$) of the typical Q1D localization length, defined
as
\begin{align}
\label{DEFLambdac}
\Lambda_{p} &= \frac{2\xi_{p}}{M}, &
\Lambda_{o} &= \frac{2\xi_{o}}{M},
\end{align}
for both types of BCs (the factor 2 in this definition is
standard convention). We verify Eq.\ (\ref{eq:ours-surface})
for (a) the metal-to-(ordinary) insulator transition in the spin-orbit
(symplectic) class
[class AII of Ref.\ \onlinecite{AltlandZirnbauer}],
(b) the LD transition between a metal and a $\mathbb{Z}_2$ topological
insulator in the ``quantum spin Hall'' (QSH) effect\cite{CommentQSHconfuseSQH}
which also belongs to the spin-orbit (symplectic) class
[class AII of Ref.\ \onlinecite{AltlandZirnbauer}],
(c) the IQH plateau transition in the unitary symmetry class
[class A of Ref.\ \onlinecite{AltlandZirnbauer}].
[The {\it bulk} relation, Eq.\ (\ref{eq:ours-bulk}),
was already verified for systems (a)-(c),
where $x_{\rho}=0$,
 in previous
work.\cite{janssen_review, dohmen96, evers01,merkt98, obuse2007, mildenberger07}]
We finally verify numerically Eqs.\
(\ref{eq:ours-bulk}) and (\ref{eq:ours-surface}) for the spin
quantum Hall transition in symmetry class C
of Ref.\ \onlinecite{AltlandZirnbauer}.
Table \ref{tab:alpha_0} summarizes the numerical results presented in
detail in Section III. Section IV presents our conclusions.

\section{Localization length and multifractality}

In this section we provide a derivation of Eqs.\
(\ref{eq:ours-bulk}) and (\ref{eq:ours-surface}). Let us begin
with a brief discussion of the underlying assumptions. We are
interested in scaling properties of the disorder average of
some physical observable [e.g., the local DOS (LDOS)] at
an LD transition point. One can recast this disorder average
into a statistical average of a properly defined operator
${\mathcal{O}}$ in a certain field theory
(e.g., a replica or supersymmetric nonlinear sigma
model).\cite{Wegner,Efetov}
The scaling properties of
${\mathcal{O}}$ at the critical point are then
controlled by the fixed point of the renormalization group (RG)
flow of the corresponding field theory. We are now ready to
state the two important assumptions we make in our derivation:\cite{CommentPrimaryOperatorAssumption}
\begin{itemize}
\item The fixed-point theory is a conformal field theory.
\item At the fixed point of the RG transformation, the
    operator ${\mathcal{O}}$ is a primary\cite{BPZ}
    field operator in the conformal field theory.
\end{itemize}

\subsection{Finite-size scaling in cylinder geometry,
and bulk exponents}

Let us consider a disordered electronic system at its critical point, confined to a disk of radius $R$
in the 2D $x$-$y$ plane, or equivalently, the complex plane with the
coordinate $z=x+iy$. We assume that all along the boundary of the disk
there is a metallic electrode attached, thus allowing for the electron
in the system to escape.\cite{BC-comment} This (absorbing) boundary
condition introduces a finite broadening $\eta$ of the single-particle
levels in the system. We assume that the broadening is of the order of
the mean level spacing in the system. This provides a regularization for
Green's functions and the LDOS as follows: 
\begin{align}
G_\pm (z, z'; E) &= \sum_n \frac{\psi_n^*(z) \psi_n(z')}{E - E_n \pm i\eta}, \\
\rho_E(z) &= \frac{i}{2\pi}[G_+(z,z;E) - G_-(z,z;E)] \nonumber \\
&= \frac{1}{\pi} \sum_n |\psi_n(z)|^2 \frac{\eta}{(E - E_n)^2 + \eta^2}.
\label{Lorentzian}
\end{align}
Here the wave functions $\psi_n(z)$ of the closed system are normalized
in the disk: $\int_{|z|\leqslant R}|\psi(z)|^2 d^2 z =1$. The integral
of the LDOS $\rho_E(z)$ over the disk gives the global DOS $\rho_E$
multiplied by the disk area $\pi R^2$. 

Statistical properties of metallic or critical wave functions at energy
$E$ are closely related to those of the LDOS.\cite{mirlin_review} In
particular, if we are interested in the scaling of the moments of such
wave functions and the moments of the LDOS, we can write symbolically 
\begin{align}\label{wavefunction DOS relation}
|\psi_E(z)|^2 \sim \frac{\rho_E(z)}{\pi R^2 \rho_E}.
\end{align}

Disorder averages of powers of the LDOS $\rho_E(z)$ (as
well as those of products of Green's functions) are represented
by expectation values of operators in the corresponding field
theory.\cite{Wegner,Efetov} We denote this by
\begin{align}\label{DOS field theory operator relation}
\overline{\left[\rho_E(z) \right]^q} \sim \big\langle \mathcal{O}_q(z) \big\rangle,
\end{align}
where the angular brackets denote the expectation value in the
field theory.
Here $\mathcal{O}_q(z)$ is the operator which
corresponds to the $q$-th moment of $\rho_E(z)$.
(We point out that here and in what follows
the power $q$ can take any real values.\cite{continuous q comment})
In view of Eq.\ (\ref{wavefunction DOS
relation}), the same operator represents moments of the wave
function $\psi_E(z)$:
\begin{align}
\label{WF field theory operator relation}
\big(R^2 \rho_E\big)^q \,\, \overline{|\psi_E(z)|^{2q}} \sim \big\langle
\mathcal{O}_q(z) \big\rangle.
\end{align}
Notice that the global DOS is self-averaging and can be pulled
out of the disorder average along with powers of the radius
$R$. The product $R^2 \rho_E \propto \delta^{-1}$, where
$\delta$ is the mean level spacing in the disk.

Now we concentrate on the wave functions and the DOS {\it at
the critical energy}, $E=E_c$, and drop the subscript $E$. The
global DOS $\rho$ may vanish at criticality in the infinite
system. In a finite system the disorder-averaged $\rho$ always
has a power-law behavior
\begin{align}
\rho \sim R^{-x_\rho},
\label{x_rho}
\end{align}
where the exponent $x_\rho$ vanishes in the standard
Wigner-Dyson classes but is known to be non-zero in other
symmetry classes. For example, at the (2D) spin quantum Hall
transition in symmetry class C,\cite{gruzberg99, beamond02,
mirlin03} the exact value is known: $x_\rho =
1/4$.\cite{gruzberg99}

We now make use of the previously stated
assumptions\cite{CommentPrimaryOperatorAssumption} of conformal
invariance and the fact that ${\cal O}_q$ is a primary\cite{BPZ}
conformal scaling operator with
the bulk scaling dimension $x_q^b$ at the LD transition. If we
choose a point $|z| \ll R$ close to the origin of the disk,
then the one-point function (the field theory expectation
value) scales as
\begin{equation}
\big\langle{\cal O}_q(z) \big\rangle \sim R^{-x_q^b}.
\label{eq:1-point}
\end{equation}
Combining this with Eqs.\ (\ref{WF field theory operator
relation}) and (\ref{x_rho}), we obtain the scaling of the
moments of the critical wave functions:
\begin{align}
\overline{|\psi(z)|^{2q}} &\sim R^{-2q - x_q^b + q x_\rho}
\label{WF moments bulk}
\end{align}
for $|z| \ll R$. Notice that the exponent of $R$ on the right
hand side should vanish at $q = 0$, and should be $-2$ at $q=1$
due to the normalization of the wave function. These conditions
determine
\begin{align}
x_0^b &= 0, & x_1^b &= x_\rho.
\end{align}

Some important details of the definition and properties of
multifractal exponents are in order here. A slightly more
detailed\cite{Halsey} (``coarse-grained'') description of
multifractal wave functions (in 2D) involves breaking the
system into little square boxes $B_i$ of size $r \times r$ labeled
by $i$. The number of these boxes $N$ scales as $N \sim
(R/r)^2$. One then calculates the probability $p_i$ for an
electron to be in the $i$-th box as
\begin{align}
p_i = \int_{B_i} |\psi(z)|^2 d^2 z,
\end{align}
and forms the so-called average generalized inverse
participation ratios
\begin{align}
\overline{P_q} = \sum_{i=1}^N \overline{p_i^q} = N  \overline{p_i^q}.
\end{align}
(We have assumed that the system is homogeneous after disorder
average.)
Equation (\ref{WF moments bulk}) implies the
scaling relation
\begin{align}
\overline{P_q} &\sim \Big(\frac{R}{r} \Big)^{-\tau_q}, &
\tau_q = 2(q-1) + x_q^b - q x_\rho,
\label{IPR scaling}
\end{align}
where the set of exponents $\tau_q$ is usually referred to as
the multifractal spectrum.

Note that the probabilities $p_i$ whose moments
enter the definition of $\overline{P_q}$ are bounded by
$0 \leqslant p_i \leqslant 1$.
This bound implies that $\overline{P_q}$ must be a non-increasing
function of $q$, since $p_i^{q_1} \geqslant p_i^{q_2}$ for $q_1 < q_2$.
Moreover, since $p^q = \exp(q \ln p)$ is convex
as a function of $q$, the same is true for $\overline{P_q}$.
Then the multifractal spectrum $\tau_q$ in Eq.\ (\ref{IPR scaling})
must be a non-decreasing, concave function of $q$.
Generally speaking, there may be a value of $q = q_f$ where $\tau_q$ has
a horizontal tangent. Then it follows that $\tau_q =
\text{const}$ for $q \geqslant q_f$. Such change in the
behavior of $\tau_q$ from an increasing function to a constant
is often referred to as ``freezing'' or ``termination'' (see
Ref.\ [\onlinecite{evers_review}] for more details). In all
known cases the value $q_f$ where such termination occurs
satisfies $q_f > 0$. Then we can safely use Eq.\ (\ref{WF
moments bulk}), and similar equations in the following
sections, in the vicinity of $q=0$ without worrying about a
possible termination transition.

Expanding both sides of Eq.\ (\ref{WF moments bulk}) in $q$
about $q=0$ yields the typical scaling exponent, Eq.\
(\ref{eq:alpha_0}), where
\begin{equation}
\alpha_0^b = 2 - x_\rho + \frac{dx_q^b}{dq}\Big|_{q = 0}.
\label{eq:alpha_0=2+dDdq}
\end{equation}

Next, let us consider the conformal mapping
\begin{align}
w &= \frac{M}{2\pi}\ln z, & z = \exp\Big( \frac{2\pi}{M} w\Big),
\label{conformal_mapping1}
\end{align}
which maps the disk to the semi-infinite cylinder of
circumference $M$ in the complex $w$-plane,
\begin{align}
w & =u+iv, & u \leqslant L \equiv \frac{M}{2\pi}\ln R, &&
0 \leqslant v < M,
\end{align}
with an absorbing boundary condition at $u = L$.
The assumption that ${\cal O}_q$ is a primary conformal
operator\cite{BPZ} allows us to relate its expectation value
on the cylinder to that in the disk:
\begin{align}
\big\langle{\cal O}_q(w) \big\rangle &= \Big| \frac{dz}{dw}
\Big|^{x_q^b}
\big\langle{\cal O}_q(z) \big\rangle \nonumber \\
&\sim
\Big( \frac{2\pi}{M} \Big)^{x_q^b} \exp\Big[ -\frac{2\pi}{M}
x_q^b (L - u)\Big].
\label{ConfMapBulk}
\end{align}
This immediately gives the moments of the LDOS in the cylinder:
\begin{align}
\overline{\left[\rho(w) \right]^q} \sim \exp\Big[ -\frac{2\pi}{M}
x_q^b (L - u)\Big].
\label{LDOS-moments-cylinder}
\end{align}
From the exponential decay\cite{LDOS-moments} of the moment
$\overline{\left[\rho(w) \right]^q}$ away from the end of the
semi-infinite cylinder in Eq.\ (\ref{LDOS-moments-cylinder}), for
sufficiently small positive values of $q$, we identify the
``$q$-dependent localization length'' $\xi_p(q)$ in the cylinder
geometry as 
\begin{equation}
\xi_p(q)=\frac{M}{2\pi x_q^b}.
\label{eq:lambda_q^p}
\end{equation}
(Here `$p$' denotes again the `periodic' BCs of the cylinder.)
The typical Q1D localization length $\xi_p$ in cylinder
geometry is read off from the typical exponential decay of
the LDOS
away from the end of the semi-infinite cylinder:
\begin{equation}
\overline{\ln \rho(w)} = -\frac{|L - u|}{\xi_p} + \ldots.
\end{equation}
Expanding again Eq.\ (\ref{LDOS-moments-cylinder}) in $q$ about
$q=0$  yields
\begin{align}
\label{lambdapCylinder}
\frac{M}{\xi_p} = 2\pi \frac{dx_q^b}{dq}\Big|_{q = 0}
= 2\pi (\alpha_0^b - 2 + x_\rho),
\end{align}
where we have used Eq.\ (\ref{eq:alpha_0=2+dDdq}). This is our
previously mentioned result, Eq.\ (\ref{eq:ours-bulk}), which
generalizes Eq.\ (\ref{eq:janssen}) to all symmetry classes,
including those with critical DOS.

In Section \ref{subsec:SQH} we numerically verify Eq.\
(\ref{eq:ours-bulk}) for the spin quantum Hall effect (symmetry
class C) by computing numerically the FSS amplitude
$\Lambda_p = 2\xi_p/M$ of the typical Q1D localization
length $\xi_p$ in cylinder geometry; according to our
above-obtained result (\ref{lambdapCylinder}) this quantity is
predicted to equal
\begin{align} \Lambda_p = \frac{1}{\pi
(\alpha_0^b - 2 + x_\rho)},
\label{eq:Lambda_BdG_PBC}
\end{align}
with $x_\rho = 1/4$.

\subsection{Finite-size scaling in strip geometry, and surface (boundary)
multifractal exponents}
\label{subsec:strip}

We now apply the same arguments to discuss finite-size scaling in the
presence of open (reflecting) BCs in the transverse direction (strip
geometry). 

For this purpose we first consider the operator ${\cal O}_q$ placed
close to the origin in the interior of the half disk 
$|z| \leqslant R, \text{ Im}\,z \geqslant 0$. The boundary of the system
on the real axis is assumed reflecting, and the rest is attached to a
metallic lead, as in the previous section. In this situation the
expectation value of ${\cal O}_q(z)$ for $|z| \ll R$ is given
by\cite{Suppress-z-Dependence} 
\begin{equation}
\big\langle{\cal O}_q(z) \big\rangle \sim R^{-x_q^s},
\label{eq:1-point-surface}
\end{equation}
where the boundary scaling dimension $x_q^s$ (the superscript
$s$ stands for ``surface'') is typically different from the
bulk dimension $x_q^b$. In analogy with Eq.\ (\ref{WF moments
bulk}) we now have, upon making again use of Eq.\ (\ref{WF
field theory operator relation}),
\begin{align}
\overline{|\psi(z)|^{2q}} &\sim R^{-2q - x_q^s + q x_\rho},
\label{WF moments surface}
\end{align}
where the same exponent $x_\rho$ (a bulk exponent)  enters
through the global DOS. Note that Eq.\ (\ref{WF moments
surface}) still implies $x_0^s = 0$, but now,  in the boundary
case, there is no restriction on $x_1^s$ (in contrast to the
bulk case: see Eq.\ (\ref{WF moments bulk}) and the subsequent
text). Also, in complete analogy to the bulk case, the exponent
of $R$ in Eq.\ (\ref{WF moments surface}) must be a monotonic
function of $q$. Upon expanding both sides of Eq.\ (\ref{WF
moments surface}) in $q$ about $q = 0$, one obtains the scaling
exponent $\alpha_0^s$ of the typical wave function amplitude
{\it at the boundary},
\begin{align}
\overline{\ln|\psi(z)|^2}\sim -\alpha_0^s\ln R,
\label{eq:alpha_0-surface}
\end{align}
where now
\begin{equation}
\alpha_0^s = 2 - x_\rho + \frac{dx_q^s}{dq}\Big|_{q = 0}.
\label{eq:alpha_0=2+dDdq-surface}
\end{equation}

Next, in order to relate this to the strip geometry, we use the
conformal transformation
\begin{align}
\label{boundary conformal transformation}
w &= \frac{M}{\pi}\ln z, & z = \exp \Big(\frac{\pi}{M}w \Big)
\end{align}
which maps the half disk to a semi-infinite strip of width $M$
in the $w$-plane:
\begin{align}
w & = u + iv, & u \leqslant L \equiv \frac{M}{\pi} \ln R, && 0 \leqslant
v \leqslant M.
\end{align}
The expectation value on the strip now follows again since
${\cal O}_q$, as a primary\cite{BPZ} conformal operator,
transforms simply under conformal transformations,
\begin{align} \big\langle{\cal O}_q(w) \big\rangle &= \Big|
\frac{dz}{dw} \Big|^{x_q^s}
\big\langle{\cal O}_q(z) \big\rangle \nonumber \\
&\sim
\Big( \frac{\pi}{M} \Big)^{x_q^s} \exp\Big[ -\frac{\pi}{M}
x_q^s (L - u)\Big].
\end{align}
From this we obtain the exponential decay of the moments of
the LDOS away from one end of the strip,
\begin{align}
\overline{\left[\rho(w) \right]^q} &\sim \exp\Big[ -\frac{\pi}{M} x_q^s
(L - u)\Big]. \label{LDOS moments in strip}
\end{align}
As in the bulk case, 
the exponential decay\cite{LDOS-moments} of the right hand side in Eq.\
(\ref{LDOS moments in strip}), for sufficiently small positive values of
$q$, gives the ``$q$-dependent Q1D localization length'' along the strip 
\begin{equation}
\xi_o(q)=\frac{M}{\pi x_q^s}.
\end{equation}
As before, the  typical Q1D localization length $\xi_o$ in strip
geometry is obtained by expanding both 
sides of Eq.\ (\ref{LDOS moments in strip}) in $q$ about $q=0$,
\begin{align}
\frac{M}{\xi_o} = \pi \frac{dx_q^s}{dq}\Big|_{q = 0}
= \pi (\alpha_0^s - 2 + x_\rho),
\end{align}
where we have used Eq.\ (\ref{eq:alpha_0=2+dDdq-surface}). This
is our previously-announced result from Eq.\
(\ref{eq:ours-surface}).

In subsequent sections we verify Eq.\ (\ref{eq:ours-surface})
for various LD transitions by computing numerically the FSS
amplitude $\Lambda_o = 2\xi_o/M$ of the Q1D  typical
correlation length $\xi_o$ on the strip (`$o$'=`open',
reflecting BCs) which, according to our result, is predicted to
be equal to
\begin{align}
\Lambda_o = \frac{2}{\pi (\alpha_0^s - 2 + x_\rho)}.
\label{eq:Lambda_OBC}
\end{align}

\section{Numerical results}

In this section we present the results of our numerical
simulations supporting Eqs.\ (\ref{eq:Lambda_BdG_PBC}) and
(\ref{eq:Lambda_OBC}). For convenience, we have gathered all
the relevant fitting parameters and other numerical data in a
single table \ref{tab:Lambda}.

\begin{table*}[t]
\caption{A list of parameters obtained or used in the FSS
 analysis for the scaling functions defined in
Eqs.\ (\ref{eq:scaling_series}) and
(\ref{eq:scaling_series_even}). Here $\Lambda_c$, $\nu$, and
$y$ are obtained through fitting. $N_d$ and $N_p$ denote the
numbers of data points and fitting parameters used in the
fitting procedure, respectively. The fitting functions are
truncated at the orders $P$ and $Q$. $\chi^2$ and $g$ denote
the the values of chi squared and the goodness of fit
probability, respectively.} \label{tab:Lambda}
\begin{tabular}{lccccccccccc}
\hline
\hline
system &BCs &scaling function &$\Lambda_c$ & $\nu$ & $y$& $N_d$
& $N_p$ & $P$ & $Q$ &$\chi^2$ & $g$
\\
\hline
symplectic (M-I) & reflecting & Eq.\ (\ref{eq:scaling_series})
&$1.50\pm0.01$ & $2.79\pm0.03$ & $-1.03\pm0.03$ & $85$ & $9$
&$2$ & $2$ & $86.2$ & $0.2$\\
symplectic (M-QSH) & reflecting& Eq.\
(\ref{eq:scaling_series})& $7.20\pm0.01$ & - & $-0.81\pm0.08$ &
$8$&$3$ &$0$ &$0$&$7.2$ &
$0.2$\\
IQH & reflecting& Eq.\ (\ref{eq:scaling_series})
& $1.624\pm0.002$ & $2.55\pm0.01$ & $-1.29\pm0.04$ &$134$ & $6$
& $3$ &	$2$ & $144.0$ & $0.2$\\
SQH in class C &periodic & Eq.\ (\ref{eq:scaling_series_even})
& $0.8189\pm0.0004$ & $1.335\pm0.016$ & $-0.94\pm0.01$ & $73$ &
$8$ & $2$ & $2$ &$56.2$ & $0.7$ \\
SQH in class C &reflecting & Eq.\ (\ref{eq:scaling_series}) &
$1.101\pm0.002$ & $1.335\pm0.005$ & $-1.05\pm0.02$ & $93$ & $9$
& $3$ & $2$ &$86.1$ & $0.4$  \\
\hline
\hline
\end{tabular}
\end{table*}

In this section we have to distinguish off-critical and critical
values of the Q1D localization lengths, $\xi$ and $\xi_c$, and
the corresponding FSS amplitudes, $\Lambda$ and $\Lambda_c$
(for both periodic and open BCs).
All $\xi$ and $\Lambda$ that have appeared in the previous sections
denoted values at the critical point.

\subsection{Spin-orbit (symplectic) symmetry  class}
\label{sec:ordinary symplectic}

To compute the localization length at the LD transition in the
symplectic class, we employed the so-called SU(2)
model,\cite{asada} a tight-binding model on the square lattice,
with random on-site disorder and fully random SU(2) hopping.

\subsubsection{Localization length (strip geometry)}

\begin{figure}
\vspace{-8mm}
\begin{center}
\includegraphics[width=8cm]{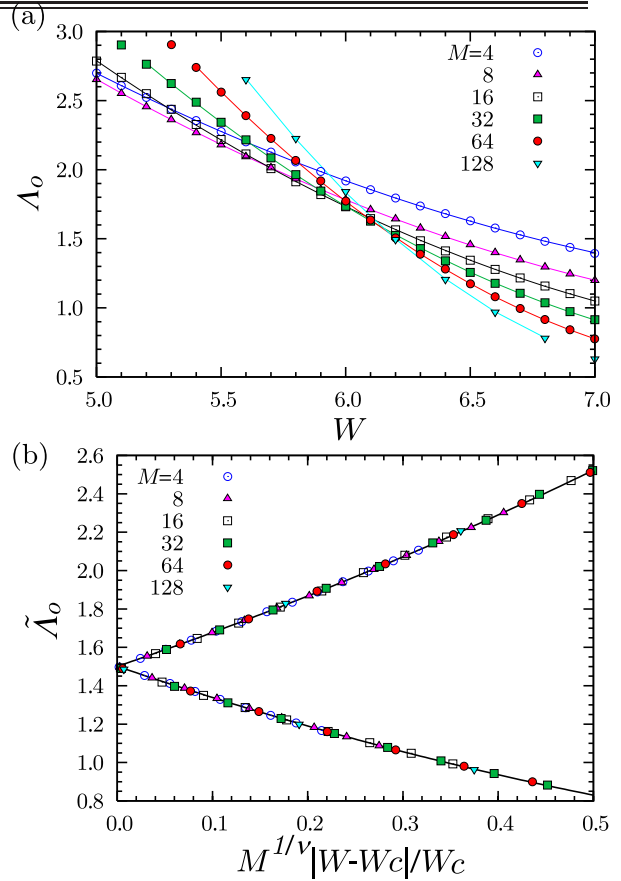}
\end{center}
\caption{(Color online)
(a) Dependence of $\Lambda_o$ on $W$ at $E=0$ for various
values of $M$. $\widetilde{\Lambda}_o$ at $E=0$. (b) Scaling
plot of $\widetilde{\Lambda}_o$ at $E=0$ -- see Eq.\
(\ref{eq:Lambda'}). The obtained parameters are
$\Lambda_{o,c}=1.50\pm0.01$, $W_c=6.192\pm0.007$,
$\nu=2.790\pm0.025$, $y=-1.026\pm0.03$, $a_1=-1.69 \pm 0.03$,
$a_2=0.70\pm0.02$, $b_0=1.24\pm0.03$, $b_1=-2.36 \pm0.08$, and
$b_2=4.7\pm0.3$. } \label{fig:SU(2)}
\end{figure}

We obtained the typical localization length from the smallest
Lyapunov exponent of transfer matrices for very long Q1D
lattices. We imposed hard-wall, i.e., reflecting BCs in the
transverse direction and hence our Q1D samples had strip
geometry. Our systems had a maximum size $M=128$ in the
transverse direction.  Figure \ref{fig:SU(2)}(a) shows the FSS
amplitude $\Lambda_{o}=2\xi_{o}/M$ of the typical Q1D
localization length as a function of the on-site disorder
strength $W$ for various system sizes $M$ and at fixed
energy $E=0$ (band center). The curves for the various system
sizes intersect at different points reflecting large
finite-size effects, in contrast to the case of periodic
BCs.\cite{asada}

To determine the critical value of the FSS amplitude
$\Lambda_{o,c}$, we performed a FSS analysis incorporating
corrections to scaling arising from the leading irrelevant
scaling variable.\cite{slevin99} Specifically, we took a
scaling function for the FSS amplitude of the form
$\Lambda=F(\chi M^{1/\nu},\zeta M^y)$, where $\chi$ is the
relevant scaling variable, and $\zeta$ is the leading
irrelevant scaling variable whose scaling exponent $y<0$. The
exponent $\nu$ characterizes the divergence of the 2D
localization length $\xi$ upon approaching the LD transition
point, $\xi \sim \chi^{-\nu}$. We expanded the scaling function
around the critical point $W=W_c$, setting $\chi=(W-W_c)/W_c$,
\begin{equation}
\Lambda_o = \Lambda_{o,c}+\sum^P_{p=1}a_p \left(\chi
M^{1/\nu}\right)^p + M^y\sum^Q_{q=0}b_q \left(\chi
M^{1/\nu}\right)^q. \label{eq:scaling_series}
\end{equation}
We fitted the numerical data to Eq.\ (\ref{eq:scaling_series})
with $P=Q=2$ by taking $W_c$, $a_p$, $b_q$, $\nu$, and $y$ as
fitting parameters. We obtained
\begin{eqnarray}
W_c=6.192\pm0.007, \qquad \Lambda_{o,c}=1.50\pm0.01,
\label{eq:SU(2)}\\
\nu=2.79\pm0.03, \qquad y=-1.03\pm0.03. \nonumber
\end{eqnarray}
The details of the fitting are summarized in Table
\ref{tab:Lambda}.

These results are in good agreement with those obtained by
Asada \textit{et al.}\ for the SU(2) model\cite{asada} with periodic
BCs: $W_c = 6.199 \pm 0.003$ and $\nu = 2.75 \pm 0.04$ at $E =
0$. The good quality of the fit can be seen from the scaling
collapse, shown in Fig.~\ref{fig:SU(2)}(b), of the data for the
corrected FSS amplitude $\tilde{\Lambda}_{o}$ defined by
\begin{equation}
\widetilde{\Lambda}_o = \Lambda_o - M^y \sum^Q_{q=0}b_q \left(\chi
M^{1/\nu}\right)^q. \label{eq:Lambda'}
\end{equation}

\subsubsection{Surface multifractal exponent $\alpha_0^s$}

In our previous publication\cite{obuse2007} we reported the
value $\alpha_0^s=2.417\pm0.002$ for the surface exponent,
which was obtained from numerical simulations on $L\times L$
lattices of system sizes up to $L=120$.
We performed averaging over more than $6\times10^4$ disorder realizations.
The lattices had periodic BC imposed in one of the two directions,
but open BC in the other direction, so our system had the geometry
of a finite cylinder.
Here we update the value for $\alpha_0^s$ reported
in our previous work.\cite{obuse2007}
We use larger system sizes up to $L = 180$, and average over
up to $10^5$ disorder realizations.

The surface exponent $\alpha_0^s$ was obtained from the system
size dependence of the wave function amplitude in the vicinity
of the boundary, according to
\begin{equation}
\big\langle \! \big\langle \ln|\psi(x)|^2 \big\rangle \! \big\rangle
\sim-\alpha_0^s\ln L+c, \label{eq:<ln|psi|>}.
\end{equation}
Here $x=\mathcal{O}(L^0)$, $L\gg1$, and $c$ is a constant of
order $L^0$. The double angular brackets represent both
ensemble average and spatial average along the boundary of
the cylinder in each disorder realization. First we tried a linear
fitting to Eq.\ (\ref{eq:<ln|psi|>}) of our numerical data for
the left hand side of Eq.\ (\ref{eq:<ln|psi|>}), using system
sizes $24 \leqslant L \leqslant 180$, with two fitting
parameters $\alpha_0^s$ and $c$. This resulted in the value
\begin{equation}
\alpha_0^s=2.4195\pm0.0013.
\end{equation}
Substitution of this value into Eq.\ (\ref{eq:Lambda_OBC})
gave
\begin{equation}
\Lambda_{o,c}=1.518\pm0.005.
\end{equation}
This analysis, however, ignored corrections from irrelevant
scaling variables and was not quite correct, since we now know
from the previous subsection that such corrections are
appreciable for the FSS amplitude $\Lambda$ for open BC. We
therefore re-analyzed the data, assuming scaling with
corrections from the leading irrelevant variable.\cite{evers01} We define
\begin{equation}
A(x):=-\frac{\big\langle \! \big\langle
\ln|\psi(x)|^2 \big\rangle \! \big\rangle}{\ln L}\sim
\alpha_0^s
+\frac{1}{\ln L}\left(c+c'L^y\right),
\label{eq:<ln|psi|>_correction}
\end{equation}
where we take $y=-1$, as suggested by Eq.\ (\ref{eq:SU(2)}).
The fitting of the same data to Eq.\ (\ref{eq:<ln|psi|>_correction})
yielded
\begin{equation}
\alpha_0^s=2.429\pm0.006,
\label{alpha_0^s_SU(2)}
\end{equation}
which leads to
\begin{equation}
\Lambda_{o,c}=1.48\pm0.02
\end{equation}
with the help of
Eq.\ (\ref{eq:Lambda_OBC}).
We see that the
$\Lambda_{o,c}$ obtained from the transfer matrix method
(\ref{eq:SU(2)}) is consistent with these results. The value of
$\alpha_0^s$ reported in Eq.\ (\ref{alpha_0^s_SU(2)}) has
larger error bars, which needs to be improved in future
numerical work.

\subsection{Metal to $\mathbb{Z}_2$ topological insulator transition
in quantum spin Hall systems}

The $\mathbb{Z}_2$ topological insulator is a time-reversal
invariant topological insulator in two dimensions, which
possesses a topologically protected Kramers pair of extended
edge states at its boundaries.\cite{Kane}  The $\mathbb{Z}_2$
topological insulating states can be realized in materials with
strong spin-orbit interactions, as evidenced by recent
experiments on HgTe/(Hg,Cd)Te quantum wells.\cite{koenig07}
In the presence of
disorder, this system undergoes a two-dimensional metal-insulator
transition from a $\mathbb{Z}_2$ topological insulator to a
metal, as one changes the Fermi energy.
On symmetry grounds, this LD transition is expected to
belong to the spin-orbit (symplectic) symmetry
class.\cite{obuse07b} Indeed, the critical exponent $\nu$ for
the diverging localization length (a bulk property) at the
metal to $\mathbb{Z}_2$ topological insulator transition  is
found to agree with the value obtained for the SU(2)
model,\cite{obuse07b} which describes the metal to (ordinary)
insulator transition in this symmetry class.
Similar agreement is found for the multifractal exponents
for critical wave functions in the bulk.\cite{obuse08} However,
the multifractal exponents characterizing wave function
amplitudes at the sample {\it boundary} turn out to be
different at the two metal-insulator transitions.

Here we show that, at the metal to $\mathbb{Z}_2$ topological
insulator transition, the FSS amplitude $\Lambda_{o,c}$ (Eq.\
(\ref{DEFLambdac})) for the typical Q1D correlation length in
strip geometry, is related by conformal invariance to
the boundary multifractal exponent $\alpha_0^s$ at the same transition.

\subsubsection{Localization length (strip geometry)}

\begin{figure}
\includegraphics[width=0.43\textwidth]{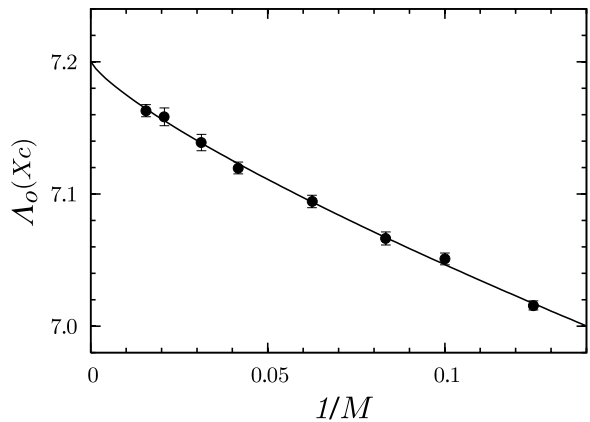}
\caption{
$M$ dependence of $\Lambda_o$ at the metal to topological
quantum spin Hall insulator transition. The solid curve is a
fit to Eq.~(\ref{eq:Lambda_c_correction}) with $\Lambda_{o,c} =
7.20 \pm 0.01$, $y = -0.81 \pm 0.08$, and $b_0 = -1.0 \pm 0.1$.
} \label{fig:Lambda_QSH}
\end{figure}

To compute the localization length at the metal to
$\mathbb{Z}_2$ topological insulator transition, we employed the
quantum spin Hall network model.\cite{obuse07b,obuse08} An
important parameter in this network model is the one
controlling the probability of tunneling at the  nodes of the
network, which we denote by $X$. The numerical results shown
below were obtained at the critical point $X_c=0.971$ with
fully random SU(2) spin rotation symmetry on each
link.\cite{obuse08}
Figure \ref{fig:Lambda_QSH} shows the dependence of
the FSS amplitude $\Lambda_o(M) := 2 \xi_o(M)/M$ of the
typical Q1D localization length $\xi_o(M)$ on a strip of
width $M$ ($M=8,10,12,16,24,32,48,64$). Here $M$ is the number
of nodes of the network model in the transverse direction
across the Q1D strip. This corresponds to transfer matrices of
size $4M \times 4M$. In order to find the critical value
$\Lambda_{o,c}$ of the FSS amplitude  $\Lambda_o$ in the large
$M$ limit, we assumed that $\Lambda_o$ at $X=X_c$ has a
power-law finite-size correction due to a leading irrelevant
variable with dimension $y < 0$:
\begin{equation}
\Lambda_o(X=X_c)=\Lambda_{o,c} + b_0 M^y.
\label{eq:Lambda_c_correction}
\end{equation}
Fitting the data to this form (see
Fig.~\ref{fig:Lambda_QSH}), we obtained
\begin{equation}
\Lambda_{o,c} = 7.20\pm0.01
\label{eq:Lambda_c_QSH}
\end{equation}
with $y=-0.81\pm0.08$ and $b_0=-1.0\pm0.1$. The details of the fitting are
summarized in Table \ref{tab:Lambda}.

\subsubsection{Surface multifractal exponent $\alpha_0^s$}

The surface multifractal exponent at the metal to
$\mathbb{Z}_2$ topological insulator transition was obtained in
Ref.~[\onlinecite{obuse08}].
By using larger system sizes this value was recently improved
in Ref.~[\onlinecite{shinsei09}] to
\begin{equation}
\alpha_0^s=2.091\pm0.002.
\label{NewEquation}
\end{equation}
Substituting the improved value
into Eq.\ (\ref{eq:Lambda_OBC})
yields the FSS amplitude
\begin{equation}
\Lambda_{o,c}=7.00\pm0.15.
\label{Lambda-7}
\end{equation}
This value is consistent with Eq.\ (\ref{eq:Lambda_c_QSH}).
The larger error bar in Eq.\ (\ref{Lambda-7}) results from the fact
that the denominator in Eq.\ (\ref{eq:Lambda_OBC}) (with $x_\rho = 0$)
contains $\alpha_0^s-2 = 0.091\pm0.002$.
Neither of the numerical analyses in
Refs.\ [\onlinecite{obuse08}, \onlinecite{shinsei09}], used to
obtain Eq.\ (\ref{NewEquation}), included effects of the
leading irrelevant variable, in contrast to Eq.\ (\ref{eq:Lambda_c_QSH}).
These effects may influence the value
of $\alpha_0^s$ and possibly result in better agreement with
Eq.\ (\ref{eq:Lambda_c_QSH}).

\subsection{Plateau transition in the integer quantum Hall effect}

To compute the localization length $\xi_o$ and the surface multifractal
exponent $\alpha_0^s$ at the plateau transition in the IQH effect, we employed
the Chalker-Coddington network model\cite{chalker88,
kramer_review} in strip geometry with $M$ nodes in the
transverse direction across the strip. This corresponds to
transfer matrices of size $2M \times 2M$.
The plateau transition is reached by tuning a parameter $\theta$
which controls the tunneling probability
at the nodes of the network model.
For this model the critical value $\theta_c$ is known exactly.

\subsubsection{Localization length  (strip geometry)}

\label{sec:IQH_Lambda}

\begin{figure}
\includegraphics[width=0.43\textwidth]{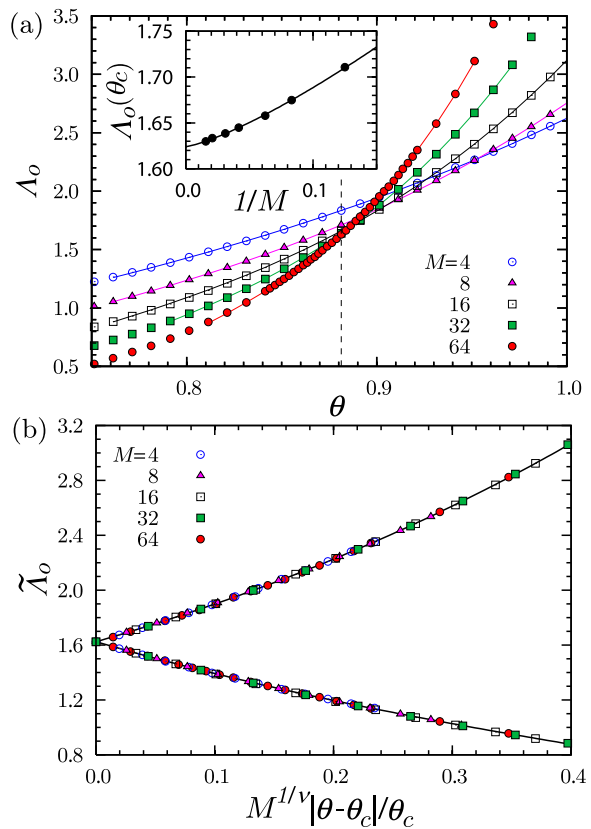}
\caption{(Color online)
(a) Dependence of $\Lambda_o$ on the node parameter $\theta$ in
the Chalker-Coddington model of the strip geometry. The
vertical dashed line indicates the critical point
$\theta = \theta_c$. Inset: $M$ dependence of
$\Lambda_o$ at $\theta=\theta_c$; the solid curve is a fit to
Eq.\ (\ref{eq:Lambda_c_correction}). (b) Scaling plot from FSS
analysis including corrections from the leading irrelevant
scaling variable. The parameters used for the plot are
$\nu=2.55\pm0.01$, $a_1=2.518\pm0.016$, $a_2=2.179\pm0.027$,
$a_3=1.393\pm0.051$,
$b_0=1.26\pm0.7$,
$b_1=2.016\pm0.086$, and
$b_2=-0.73\pm0.26$. } \label{fig:Lambda_IQH}
\end{figure}

The typical  localization length $\xi_o$ in Q1D strip
geometry was computed numerically from the smallest Lyapunov
exponent of the transfer matrices. The largest system size (the
number of network model nodes in the transverse direction) that we
studied, was $M=64$. Figure \ref{fig:Lambda_IQH}(a) shows the
FSS amplitude $\Lambda_o = 2\xi_o/M$ of the typical
localization length as a function of the network model
tunneling parameter $\theta$ for various transverse system
sizes $M$.\cite{chalker88} For $\theta>\theta_c$ the network
model is in the quantum Hall phase.\cite{comment-IQH phase}
As seen from Fig. \ref{fig:Lambda_IQH}, the crossing point of
the curves moves towards $\theta=\theta_c$ as
$M$ increases, indicating the presence of finite-size
corrections. To find the critical value of the FSS amplitude
$\Lambda_o$ of the typical Q1D correlation length in the large
$M$ limit, we fitted the data to Eq.\
(\ref{eq:Lambda_c_correction}) [see the inset of
Fig.~\ref{fig:Lambda_IQH}(a)], to obtain
\begin{equation}
\Lambda_{o,c} = 1.624 \pm 0.002,
\label{eq:Lambda_c^o-IQH}
\end{equation}
$y = -1.29 \pm 0.04$, and $b_0 = 1.26 \pm 0.7$. The details of
the fitting are summarized in Table \ref{tab:Lambda}. Figure
\ref{fig:Lambda_IQH}(b) shows the data collapse from the FSS
analysis using Eqs.\ (\ref{eq:scaling_series}) and (\ref{eq:Lambda'})
with $\chi=(\theta-\theta_c)/\theta_c$ and the values of
$\Lambda_{o,c}$, $y$, and $b_0$ obtained above.
This FSS analysis also yielded $\nu=2.55\pm0.01$ for the critical
exponent of the diverging
(2D bulk) localization length, which is close to the value
obtained in a recent large-scale numerical study, $\nu =
2.593 \pm 0.006$.\cite{SlevinOhtsuki09}

\subsubsection{Surface multifractal exponent $\alpha_0^s$}
\label{subsec:IQH-alpha-0-s}

The surface multifractal exponent $\alpha_0^s$ at the plateau
transition was recently obtained by the present
authors\cite{obuse08b} and by Evers, Mildenberger, and
Mirlin.\cite{evers08} It was found in these works that the
multifractal analysis for the Chalker-Coddington model suffers
from large finite-size corrections. To reduce these
corrections, we have used, in the multifractal scaling analysis
in Ref.\ [\onlinecite{obuse08b}], numerical data obtained only
for large system sizes. Here we used an alternative approach by
taking into account corrections to scaling arising from a
leading irrelevant scaling variable using Eq.\
(\ref{eq:<ln|psi|>_correction}).

\begin{figure}
\includegraphics[width=0.3\textwidth]{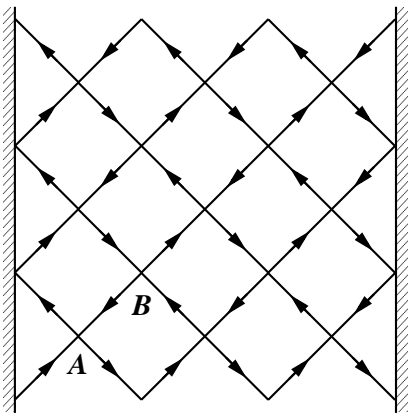}
\caption{
The Chalker-Coddington network model on a cylinder. In the
notation of section \ref{subsec:IQH-alpha-0-s} $L=3$ on this
figure. There are $4L^2 = 36$ links, and the unitary
evolution operator $U$ is a $36 \times 36$ matrix.}
\label{fig:CC-network-square}
\end{figure}
The geometry of the Chalker-Coddington network model that we
used is shown in Fig.\ \ref{fig:CC-network-square}. There are
two types of nodes forming two sub-lattices (denoted $A$ and
$B$ in the figure), such that the $A$ sublattice has the size
$L \times L$ ($L=3$ in the figure). The links of the network
form zigzag shaped rows and columns; there are $2L$ such rows
and $2L$ such columns so that the total number of links is
$4L^2$. Integer $x$ and $y$ coordinates are assigned to the
centers of links. We imposed periodic BC in the vertical $y$
direction, and reflecting BC in the horizontal $x$ direction.
The links in the first and the last columns at $x = 1$ and $x =
2L$ are called the edge links. The discrete time evolution of
wave functions defined on links of the network model is
governed by a unitary evolution operator $U$ for one discrete
time step, which is determined by the scattering $S$ matrices
at the nodes of the network model.\cite{klesse95} In our case
this operator is a $4L^2 \times 4L^2$ unitary matrix. For each
disorder realization, we obtained one critical wave function
that is the eigenvector of $U$ at $\theta=\theta_c$ and whose
eigenvalue is closest to unity among all the eigenvectors. The
largest system size we studied was $L=180$, and the
disorder average was taken over $3\times10^5$ realizations for
$L \leqslant 60$, over $5\times10^5$ realizations for $L=80$,
and over $2\times10^5$ realizations for $L=120,180$.

\begin{figure}[t]
\includegraphics[width=0.45\textwidth]{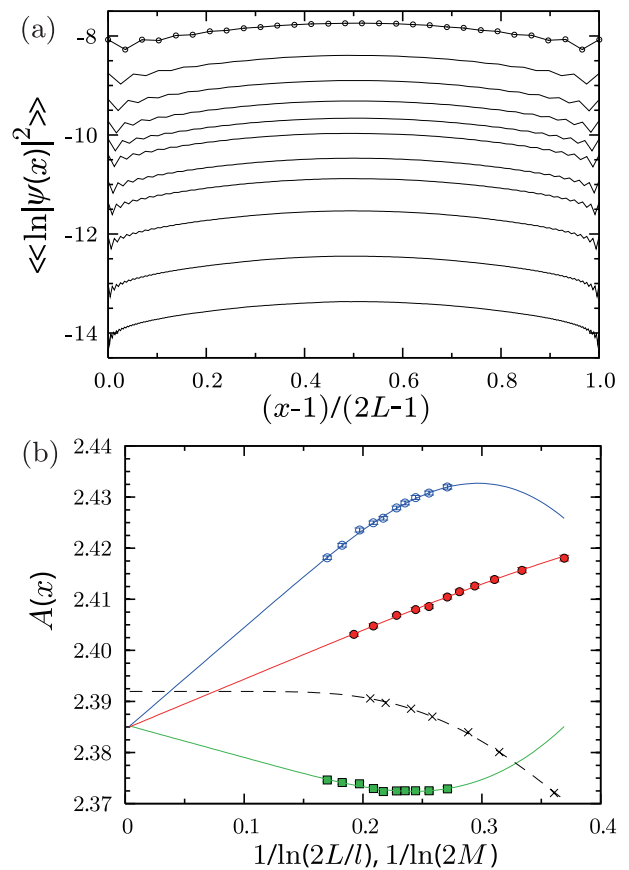}
\caption{
(Color online) (a) Spatial dependence of logarithm of
probability density $\langle \! \langle \ln |\psi(x)|^2 \rangle
\! \rangle$. The system size is changed as $L = 15, 20, 25, 30,
35, 40, 50, 60, 80, 120, 180$ from the top to the bottom. (b)
Dependence of $A(x) := - \langle \! \langle\ln|\psi(x)|^2
\rangle \! \rangle/\ln L$ on the effective system size $L/l$ at
$x=1$ (squares, $l=1$), 2 (open circles, $l=1$) and
at edge plaquettes with coarse-graining (filled circles,
$l=2$), where $l$ is the (linear) size of boxes  used to define
the coarse-grained wave function probabilities. The solid
curves are the fits to Eq.\ (\ref{eq:<ln|psi|>_correction}).
Also shown by crosses is the dependence of $2[1 + 1/\pi
\Lambda(\theta = \theta_c)]$ on the width $M$ and its fit
(dashed curve) to Eq.\ (\ref{eq:Lambda_c_correction}). }
\label{fig:WFA}
\end{figure}

Figure \ref{fig:WFA}(a) shows the $x$ dependence of
$\langle\! \langle\ln|\psi(x)|^2 \rangle \! \rangle$, where the
double angular brackets stand for both the average over
disorder realizations and the spatial average along the
periodic $y$ direction. We clearly observe in
Fig.~\ref{fig:WFA}(a) Friedel-like oscillations near the edges
of the cylinder, which become less pronounced as $L$ is
increased. (Such oscillations are absent in the SU(2) model
discussed in the previous section.) Figure~\ref{fig:WFA}(b)
shows how
$A(x) = - \langle \! \langle \ln|\psi(x)|^2 \rangle\!\rangle/\ln L$
approaches a constant value with increasing
$L$ at the left boundary ($x=1,2$).
The solid curves show the fitting of $A(x)$ to
Eq.\ (\ref{eq:<ln|psi|>_correction}) at $x=1$ (squares)
and $x=2$ (open circles). To minimize the corrections
coming from the Friedel-like oscillations, we defined the
coarse-grained wave function amplitude on each plaquette and
calculated the corresponding $A$ for the plaquettes along the
edge (shown as red filled circles). Fitting this coarse-grained data
to Eq.\ (\ref{eq:<ln|psi|>_correction}) with $y=-1.29$
obtained in Sec.\ \ref{sec:IQH_Lambda} yielded
\begin{equation}
\alpha_0^s=2.385\pm0.003,
\label{eq:alpha_0^s}
\end{equation}
where the error bars reflect only statistical errors. This
result is consistent with that of Ref.\ [\onlinecite{obuse08b}]
($\alpha_0^s=2.386\pm0.004$). Figure \ref{fig:WFA}(b) shows
that fitting of $A(x=1)$ and $A(x=2)$ gives similar values
of $\alpha_0^s$. Substituting
Eq.~(\ref{eq:alpha_0^s}) into
Eq.\ (\ref{eq:Lambda_OBC}) yields
\begin{equation}
\Lambda_{o,c}=1.654\pm0.013,
\label{Lambda_c^o-IQH}
\end{equation}
which should be compared with $\Lambda_{o,c}=1.624\pm0.002$
[Eq.\ (\ref{eq:Lambda_c^o-IQH})] obtained from the transfer matrix
calculation. As we see in Fig.~\ref{fig:WFA}(b), finite-size
corrections to $A$ and $\Lambda_o$ are still quite large at
$L=180$. This makes the extrapolation of these quantities to
$L\to\infty$ difficult; we cannot exclude the possibility of
having systematic errors in addition to the statistical errors
included in Eqs.\ (\ref{eq:Lambda_c^o-IQH}) and
(\ref{Lambda_c^o-IQH}). Given the presence of this uncertainty,
we conclude that our numerical results are consistent with
Eqs.\ (\ref{eq:ours-surface}) and (\ref{eq:Lambda_OBC}).

\subsection{Spin quantum Hall plateau transition of BdG quasiparticles in
symmetry class C} \label{subsec:SQH}

In this section we discuss the verification of Eqs.\ (\ref{eq:ours-bulk})
and (\ref{eq:ours-surface}) for symmetry class C, which is known to
possess a vanishing critical DOS ($x_{\rho}>0$). In our
simulations we used an appropriate generalization of the
Chalker-Coddington network model,\cite{kagalovsky99} which we refer to
as the class C network model.
This model has a control parameter $\epsilon$ (in the notation of
Ref.~[\onlinecite{kagalovsky99}]), and is
critical at $\epsilon=0$.
Exact values for critical exponents, $\nu=4/3$ and $x_\rho=1/4$,
were obtained through mapping to classical
percolation.\cite{gruzberg99,beamond02}
The exact values of the
bulk\cite{mirlin03} and surface\cite{subramaniam} multifractal
wave function exponents $x_q^{b,s}$ are also known at $q=2,3$.
However, exact results for the FSS amplitudes of the  typical Q1D
correlation lengths, $\Lambda_{p,c}$ and $\Lambda_{o,c}$,
and the typical wave function scaling exponents $\alpha^{b,s}_0$ are
not available.

\subsubsection{Localization length (cylinder and strip geometries)}

\begin{figure}[t]
\includegraphics[width=0.43\textwidth]{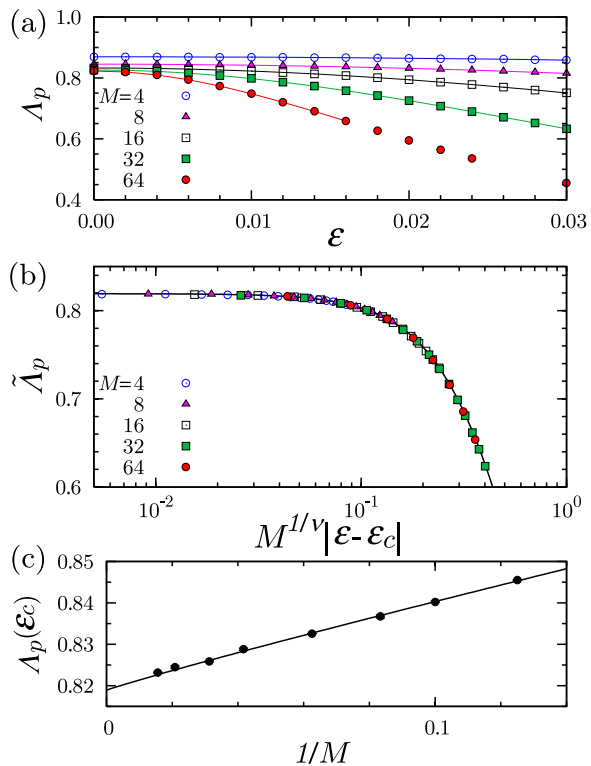}
\caption{(Color online)
(a) Dependence of $\Lambda_p$ on $\epsilon$ for several values
of $M$ in the class C network model of cylinder geometry. A
critical point is known to be located at $\epsilon_c=0$. (b)
Scaling plot of $\widetilde{\Lambda}_p$, obtained after
subtracting corrections to scaling from a leading irrelevant
scaling variable. The parameters used for the plot are $
\Lambda_{p,c} = 0.8189 \pm 0.0004$, $\nu = 1.335 \pm 0.016$, $y =
-0.94 \pm 0.01$, $a_2 = -1.66 \pm 0.10$, $a_4 = 3.64 \pm 0.33$,
$b_0 = 0.185 \pm 0.003$, and $b_2 = 0.58 \pm 0.31$. (c) $M$
dependence of $\Lambda_p$ at $\epsilon = 0$. The solid curve is
a fit to Eq.~(\ref{eq:Lambda_c_correction}). }
\label{fig:Lambda_classC_PBC}
\end{figure}

We numerically obtained the FSS amplitudes of the typical Q1D
localization length of the class C network model for both cylinder
and strip geometries. A previous numerical study\cite{kagalovsky99}
of FSS of the typical localization length in cylinder geometry
did not report the value of $\Lambda_{p,c}$.
Here we present results for the FSS amplitudes $\Lambda_{p,c}$
and $\Lambda_{o,c}$ corresponding to cylinder and strip geometries,
respectively.

{\it Cylinder Geometry}: Figure \ref{fig:Lambda_classC_PBC}(a)
shows the dependence of the FSS amplitude $\Lambda_{p}$ of the
typical Q1D correlation length on the parameter
$\epsilon$ for various values of the transverse width $M$,
obtained in cylinder geometry. The FSS amplitude
$\Lambda_p$ is symmetric about the critical point
$\epsilon_c=0$ when periodic BCs are imposed. Hence in the FSS
analysis we have to use an expansion in even powers of $\epsilon$,
\begin{equation}
\Lambda_p = \Lambda_{p,c} + \sum^{P}_{p=1}a_{2p}
\left(\epsilon M^{1/\nu}\right)^{2p}
+ M^y \sum^{Q}_{q=0}b_{2q}
\left(\epsilon M^{1/\nu}\right)^{2q}.
\label{eq:scaling_series_even}
\end{equation}
The result of fitting of the data in
Fig.~\ref{fig:Lambda_classC_PBC}(a) to Eq.\
(\ref{eq:scaling_series_even}) is shown in Fig.\
\ref{fig:Lambda_classC_PBC}(b). The dependence of $\Lambda_{p}(\epsilon_c)$
on the width $M$ at the critical point $\epsilon_c=0$ is
plotted in Fig.\ \ref{fig:Lambda_classC_PBC}(c). We obtained
\begin{equation}
 \Lambda_{p,c}=0.8189\pm0.0004
\label{eq:Lambda_classC_PBC}
\end{equation}
and $\nu=1.335\pm0.016$. The details of the fitting are
summarized in Table \ref{tab:Lambda}. The latter result is
consistent with the exact value $\nu=4/3$, indicating good
accuracy of our numerical results.

{\it Strip Geometry}:
\begin{figure}[t]
\includegraphics[width=0.43\textwidth]{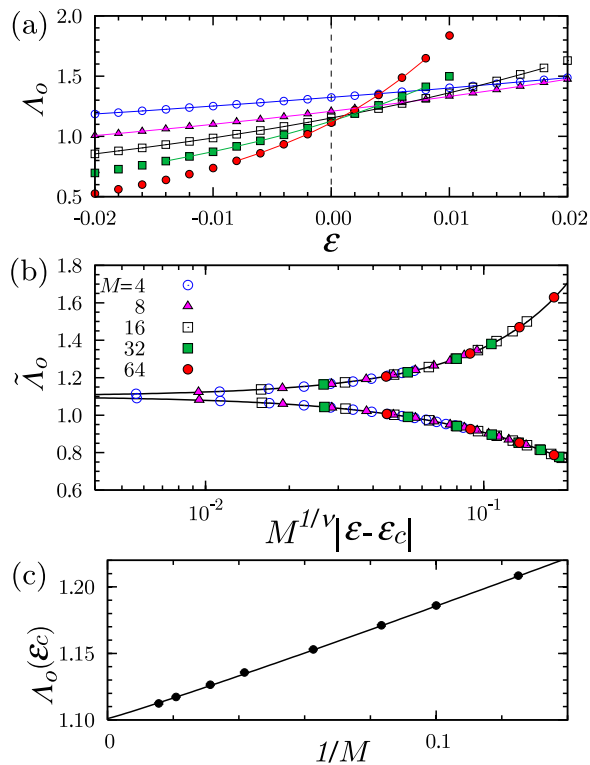}
\caption{(Color online)
(a) Dependence of $\Lambda_o$ on $\epsilon$ for several values
of $M$ in the class C network model of strip geometry. The
critical point is located at $\epsilon_c=0$. (b) Scaling plot
of $\widetilde{\Lambda}_o$ from the FSS analysis with
subtraction of corrections from a leading irrelevant scaling
variable. The parameters used for the plot are $
\Lambda_{o,c}=1.101\pm0.002$, $\nu=1.335\pm0.005$, $y = -1.05 \pm
0.02$, $a_1 = 2.225 \pm 0.024$, $a_2 = 3.221 \pm 0.083$, $a_3 =
2.91 \pm 0.20$ $b_0 = 0.960 \pm 0.015$, $b_1 = 1.846 \pm
0.095$, and $b_2 = 2.26 \pm 0.54$. (c) The $M$ dependence of
$\Lambda_o$ at $\epsilon=0$. The solid curve is a fit to
Eq.~(\ref{eq:Lambda_c_correction}).}
\label{fig:Lambda_classC_OBC}
\end{figure}
Figure \ref{fig:Lambda_classC_OBC}(a) shows the FSS amplitude
of the typical Q1D correlation length in strip geometry. With
reflecting BCs imposed in the transverse direction,
the model possesses edge states for $\epsilon>0$ (the spin quantum Hall
phase, possessing topological order).\cite{comment-IQH phase}
Since $\Lambda_o$ is
not a symmetric function of $\epsilon$, we use the FSS function
in Eq.\ (\ref{eq:scaling_series}). Figures
\ref{fig:Lambda_classC_OBC}(b) and (c) show the result of the
FSS analysis and the width $M$ dependence of the FSS amplitude
$\Lambda_{o}(\epsilon_c)$ for the typical correlation length in the strip,
respectively. From this analysis we obtained
\begin{equation}
\Lambda_{o,c}=1.101\pm0.002,
\label{eq:Lambda_classC_OBC}
\end{equation}
and $\nu=1.335\pm0.005$. The details of the fitting are
summarized in Table \ref{tab:Lambda}.

\subsubsection{Multifractal exponent $\alpha_0$}

The bulk and surface multifractal exponents $\alpha_0^b$ and
$\alpha_0^s$ for the class C network model have been obtained numerically in
Refs.~[\onlinecite{subramaniam}, \onlinecite{evers03}]:
\begin{equation}
\alpha_0^b \simeq 2.137, \qquad \alpha_0^s \simeq 2.326.
\end{equation}
Substitution of these values into
Eqs.~(\ref{eq:Lambda_BdG_PBC}) and (\ref{eq:Lambda_OBC}),
respectively, with $x_\rho=1/4$ yields
\begin{equation}
\Lambda_{p,c}=0.8225, \qquad \Lambda_{o,c}=1.105.
\end{equation}
These values are consistent with the values presented in
Eqs.~(\ref{eq:Lambda_classC_PBC}) and
(\ref{eq:Lambda_classC_OBC}) obtained by our FSS analysis.

\section{Conclusions}

In this paper we have generalized the formula relating the
multifractal exponent $\alpha_0$ of the typical wave function
amplitude in a 2D sample to the FSS amplitude $\Lambda_c$ of
the typical localization length in a Q1D sample. Our
generalization is twofold, resulting in Eqs.\ (\ref{eq:ours-bulk})
and (\ref{eq:ours-surface}). Our Eq.\ (\ref{eq:ours-bulk}) extends
the relation  to
unconventional symmetry classes where the global density of
states vanishes at criticality.
Our Eq.\ (\ref{eq:ours-surface}) extends the relation to the case
when the Q1D sample has strip geometry, instead of cylinder geometry
which was always considered in earlier studies.
In this case the multifractal exponent $\alpha_0^s$
describes the scaling of typical wave function amplitude near the
sample boundary.

We have verified generalized Eqs.\ (\ref{eq:ours-bulk})
and (\ref{eq:ours-surface}) numerically for systems in four different
universality classes: (a) the
metal-to-insulator transition in the spin-orbit (symplectic)
symmetry class, (b) the metal-to-($\mathbb{Z}_2$ topological
insulator) transition also in the spin-orbit (symplectic)
class, (c) the integer quantum Hall plateau transition,
and (d) the spin quantum Hall plateau transition.
Our numerical results are summarized in
Tables \ref{tab:alpha_0} and \ref{tab:Lambda}.

\acknowledgments

We acknowledge helpful discussions with A. Mirlin, C. Mudry and S. Ryu.
This work was partly supported by the Next Generation Super
Computing Project, Nanoscience Program from MEXT, Japan.
Numerical calculations were performed on the RIKEN Super
Combined Cluster System. H.O.\ is supported by JSPS Research
Fellowships for Young Scientists. The work of A.F.\ was
supported by a Grant-in-Aid for Scientific Research
from MEXT and JSPS, Japan
(No.~16GS0219, No.~21540332). I.A.G.\ was partially supported
by NSF Grant No.~DMR-0448820 and NSF MRSEC Grant
No.~DMR-0213745. The work of A.W.W.L.\ was supported in part by
NSF Grant No.\ DMR-0706140.


\begin{thebibliography}{50}
\bibitem{ExplainPrimary}
At any conventional non-random  second order phase transition
in two dimensions, all  scalar scaling operators which possess
scaling dimensions smaller than two are examples of conformal
primary operators.
\bibitem{CommentQSHconfuseSQH} Not to  be confused with the
    ``spin quantum Hall'' transition in symmetry class C,
    mentioned above.
\bibitem{BC-comment} Note that this boundary condition has nothing to do with the open (reflecting) boundary condition we consider in section \ref{subsec:strip}.
\bibitem{continuous q comment} In a supersymmetric sigma model
arbitrary real moments of the local density of states can be obtained
by exploiting the non-compact sector of the theory.
\bibitem{LDOS-moments}
It is known\cite{evers_review} that the exponents $x_q^b$ are positive
only within some range $0 < q < q_*$. Outside this range $x_q^b
< 0$, which leads to LDOS moments $\overline{\left[\rho(w)
\right]^q}$ that are exponentially growing away from the
cylinder boundary. This can be explained by the following
argument. In cylinder geometry, the wave functions are
localized, and the level broadening $\eta$ created by the
metallic lead attached at the edge $u = L$ is strongly
coordinate dependent. Indeed, the overlap of a wave function
localized at a point $w$ away from the cylinder edge with the
extended wave functions in the lead is exponentially small,
which results in an escape rate, that is, a level broadening
that is exponentially decaying away from the edge: 
$\eta(u) \propto e^{-(L-u)/\xi}.$
Here $\xi$ is the typical localization length. Thus, away from the edge,
the profile of the LDOS (as a function of energy) consists of
well resolved narrow Lorentzian peaks (see Eq.\
(\ref{Lorentzian})) of width of order $\eta(u)$ and height
proportional to $1/\eta(u)$. Only within the range $0 < q < q_*$
corresponding to $x_q^b > 0$ are the moments of the LDOS determined by
the minima of the LDOS profile (where the LDOS is proportional
to $\eta(u)$). For other values of $q$ the dominant contribution
to the moments of LDOS comes from the maxima in the LDOS
profile. In the described situation the behavior of the moments
of the wave functions is very different from that of the moments
of the LDOS. However, we can directly relate the latter to the
transmission through the cylinder from the metallic
lead. Essentially, the transmission will be exponentially small
away from the edge of the cylinder if the energy of the incoming
wave (in the lead) corresponds to a minimum in the LDOS profile,
but will be of order one (it cannot be bigger) if the energy is
in resonance with one of the broadened energy levels (that is, a
peak in the LDOS profile).
\bibitem{Suppress-z-Dependence} The dependence on $z$, which
    encodes the distance from the boundary, is suppressed here.
\bibitem{comment-IQH phase} In this phase $\Lambda_o$
increases with $M$, due to the presence of edge
states, which increase the conductance
(similar to the case of a metallic phase).
\end{thebibliography}
\end{document}